\documentclass{sigcomm-alternate}
\usepackage{amsfonts,latexsym,amsmath,amstext,amssymb,verbatim,epsfig,psfrag}
\usepackage{colordvi,pstcol}
\usepackage{graphicx,psboxit}
\usepackage{psfrag,graphicx,balance}

\begin{document}
\title{D-iteration: Evaluation of a Dynamic Partition Strategy}
\numberofauthors{1}
\author{
   \alignauthor Dohy Hong\\
   \affaddr{Alcatel-Lucent Bell Labs}\\
   \affaddr{Route de Villejust}\\
   \affaddr{91620 Nozay, France}\\
   \email{dohy.hong@alcatel-lucent.com}
}

\date{\today}
\maketitle

\begin{abstract}
The aim of this paper is to present a first evaluation of a dynamic partition strategy associated to the recently proposed asynchronous distributed computation scheme based on the D-iteration approach. The D-iteration is a fluid diffusion point of view based iteration method to solve numerically linear equations. Using a simple static partition strategy, it has been shown that, when the computation is distributed over K virtual machines (PIDs), the memory size to be handled by each virtual machine decreases linearly with K and the computation speed increases almost linearly with K with a slope becoming closer to one when the number N of linear equations to be solved increases.
Here, we want to evaluate how further those results can be improved when a simple dynamic partition strategy is deployed and to show that the dynamic partition strategy allows one to control and equalize the computation load between PIDs without any deep analysis of the matrix or of the underlying graph structure.
\end{abstract}

\begin{psfrags}
\section{Introduction}\label{sec:intro}
Solving efficiently a very large linear equation systems (and the related initial problems) is a very classical problem
and challenge for the algorithm design. 
The complexity of the problem to solve numerically a very large linear systems may increase rapidly with 
the dimension of the vector space.
There are many known approaches to solve such a class of problems: Gauss elimination, Jacobi iteration, Gauss-Seidel
iteration, SOR (successive over-relaxation), Richardson, Krylov, Gradient method, power iteration, QR algorithm 
etc \cite{Golub1996}, \cite{Saad}, \cite{Bagnara95aunified}, \cite{francis}, \cite{kub}, \cite{mises}.
And there are more specific approaches in more particular cases when the linear equations are associated
to a sparse matrix (in particular, in the context of PageRank equation \cite{deep}, \cite{bian}, \cite{Boldi2009}, \cite{Arasu02pagerankcomputation}: 
power method \cite{page} 
with adaptation \cite{kamvar} or extrapolation \cite{haveliwala}, \cite{kamvar2}, \cite{brezinski},
iterative aggregation/disaggregation method \cite{lang2}, \cite{kirkland}, \cite{marek}, 
adaptive on-line method \cite{serge}, etc). 
The case of the symmetric and diagonally dominant (SDD) systems 
\cite{Bunch:1989:SSA:75554.75560}, \cite{DBLP:journals/corr/cs-DS-0310036}, \cite{Koutis10approachingoptimality}
is also a very interesting case that was deeply investigated.
In parallel, there have been a lot of researches concerning the distributed computation of the
linear equations \cite{Bertsekas:1989:PDC:59912}, \cite{jela}, \cite{Lubachevsky:1986:CAA:4904.4801}, 
\cite{DBLP:journals/corr/abs-cs-0606047}, \cite{Kohlschutter06efficientparallel}, with a particular interest on asynchronous iteration scheme.

The algorithm proposed here is a new solution for a class of problem we could call diagonally dominant (DD) systems
based on the recent research results on the D-iteration. 
The D-iteration method was initially introduced in \cite{dohy} to solve numerically the eigenvector of the PageRank
equation (the eigenvector defining the score of the page importance).
Its applicability in a general linear equation has been described in \cite{d-algo}.
The distributed architecture based on this algorithm was first proposed in \cite{distributed} and
then evaluated through simulations in \cite{dist-test} when static partition strategies are applied. 
It has been shown in \cite{dist-test} that, 
when the computation is distributed over $K$ virtual machines (PIDs), 
the memory size to be handled by each virtual machine decreases linearly with $K$ and the computation 
speed increases almost linearly with $K$ with a slope becoming closer to one when the number $N$ of 
linear equations to be solved increases. However, those results were obtained under the assumption that
the information diffusion cost can be neglected, in particular the computation cost of the fluid
quantities to be diffused were neglected. Such an assumption is not realistic when $K$ becomes larger
or more precisely when $N/K$ becomes smaller.

Refining and redesigning the algorithms that were proposed in \cite{d-algo, distributed, dist-test, update},
we propose here to revisit the results in \cite{dist-test} and 
evaluate the benefit of a simple and natural dynamic partition strategy in order to
control and equalize the work load of each virtual machine when parallel computation is used.
Such a dynamic scheme may be also required when we assume that the underlying graph structure
is evolving continuously in time and updates are applied continuously (cf. \cite{update}).

In Section \ref{sec:d-archi}, we describe the distributed architecture that is
considered in this paper. Section \ref{sec:eval} presents the evaluation analysis
based on synthetic data and dataset of web graph.

\section{Distributed architecture}\label{sec:d-archi}
In this paper, we will evaluate the performance of the proposed distributed algorithm
focusing to the eigenvector problem associated to PageRank type equation. However, the algorithm is
described here in a more general case.
We assume given a square matrix $P$ of size $N\times N$ and an initial condition $B$ (a vector of size $N$).
The D-iteration applied on $(P, B)$ solves $X$ (a vector of size $N$) satisfying:
$$
X = P.X + B.
$$
The approach proposed here should work as soon as the spectral radius of $P$ is strictly less than 1
(this is what we could call a diagonally dominant system that was mentioned in the introduction).
In particular, the entries of $P$ or $B$ may be positive or negative (cf. \cite{d-algo}).
However, for a better intuitive understanding, we chose here to focus on the case where all
entries of $P$ are non-negative and implicitly associated to a transition matrix.

\subsection{D-iteration: diffusion approach}\label{sec:da}

We recall that the D-iteration is based on the fluid diffusion approach where 
one step of the iteration consists in choosing a node $i_n$ ($n$-th step) and
diffusing all fluid at node $i_n$ to its children nodes (non zero entries of the $i_n$-th column
of $P$): at each step of the iteration, we keep two state vectors: the
current residual/transient fluids are described by the vector $F_n$ and the history 
(counting the amount of diffused fluid by each node) of the fluid diffusion by the vector $H_n$.

Below, an adaptation of the pseudo-code in \cite{dist-test} for the general case:
\begin{verbatim}
Initialization:
  For i=1..N:
    H[i] := 0;   // History (counter)
    F[i] := B_i; // Fluid

Iteration:
While ( r > Target_Error )
  Choose i;    // node selection
  sent := F[i];
  H[i] += sent;
  F[i] := 0;
  For (j such that p(j,i) != 0):
    F[j] += sent * p(j,i);
  r := |F| = sum_j |F[j]|;
\end{verbatim}

When the above scheme converges (DD system), we have asymptotically (when \verb+Target_Error+ $\to$ zero) $X = H$.

\subsection{Distributed algorithm}\label{sec:distalgo}
We assume that the set $\Omega = \{1,..,N\}$ is partitioned in
$K$ sets $\Omega_k$, $k=1,..,K$ (static or dynamic, see Section \ref{sec:partition}). 
We set $L$ the number of non zero entries of the matrix $P$ (total number of
links).

\subsubsection{Local information and diffusion}
We distribute the computation tasks of the D-iteration scheme between $K$
virtual machines (we will call PIDs) as follows (cf. \cite{dist-test}):

\begin{itemize}
\item each $PID_k$ keeps information on: 
  \begin{itemize}
  \item the set of nodes it is responsible for: $\Omega_k$; 
  \item the extracted matrix $C_k(P) = (p_{ij})_{i\in\Omega, j\in\Omega_k}$, the column vectors of $P$ corresponding to 
      $\Omega_k$;
  \item the marginal fluid vector $[F]_k = ((F)_i){i\in\Omega_k}$;
  \item the marginal history vector $[H]_k = ((H)_i){i\in\Omega_k}$;
  \item the previous history vector $[H_{old}]_k = ((H)_i){i\in\Omega_k}$, the history vector value
       at the moment of the last fluid transmission (to other PIDs);
  \item its activity state: {\em active} or {\em idle} state;
  \item the target error value: {\em target\_error};
  \end{itemize}
\item each $PID_k$ maintains two local variables (evaluated periodically):
  \begin{itemize}
  \item the local residual fluid: $r_k = |[F]_k|$;
  \item the fluid to be transmitted: $s_k = |C_k(P) ([H]_k - [H_{old}]_k)|$;
  \end{itemize}
\item each $PID_k$ applies the local diffusion algorithm (*) below (when not in idle state);
\item activity state:
  \begin{itemize}
    \item initialized to active;
    \item $PID_k$'s state is set to idle when
      $$r_k < \max(s_k/10.0, target\_error\times\epsilon/K/10),$$ where $\epsilon$ is a factor
      depending on $P$: for PageRank equation, $\epsilon = 1-damping\_factor$;
  \end{itemize}

\item each $PID_k$ select the node to be diffused by a cyclic check-up of elements of $\Omega_k$
  of the condition: $$(F)_i\times w_i > T_k,$$ where $T_k$ is a threshold value initialized to 
  an arbitrary value larger than $\max_{i\in\Omega_k}(F)_i\times w_i$ and $w_i$ the weight we associate
  to the node $i$; the greedy approach would set $w_i = 1$; other candidates are: 
  $w_i = 1/(\#out_i)$ or $w_i = 1/(\#out_i\times \#in_i)$, where $\#out_i$ and $\#in_i$ are respectively
  the number of of the outgoing links from (number of non zero entries of $i$-th column of $P$)
  and the incoming links to node $i$ (number of non zero entries of $i$-th line of $P$). By default,
  we choose in this paper $w_i = 1/(\#out_i)$.
  When for all $i$, the condition $(F)_i \times w_i > T_k$ is not satisfied, we apply: $T_k := T_k/\gamma$
  (by default, $\gamma=1.2$).
\end{itemize}

\begin{verbatim}
Local diffusion for PID(k): (*)
  Choose i in Omega_k;
  sent := F[i];
  H[i] += sent;
  F[i] := 0;
  For ( j in Omega_k such that p(j,i) != 0 ):
    F[j] += sent * p(j,i);
\end{verbatim}

\subsubsection{Fluid exchange}
The transmission of fluid from $PID_k$ to other PIDs is done when:
\begin{eqnarray}\label{eq:send}
s_k &>& r_k/2.
\end{eqnarray}
The idea is just to anticipate a bit the moment when $s_k$ and $r_k$ becomes equal.
The PIDs ($PID_{k'}$) receiving $received = |[C_k(P) ([H]_k - [H_{old}]_k)]_{k'}|$ fluids reinitialize $T_{k'}$ to 
$\min(T_{k'} \times (r_{k'}+received)/r_{k'}, received)$.

\subsection{PID modelling}\label{sec:pid}
As in \cite{dist-test},
we consider a time stepped approximation for the simulation of the distributed computation cost
(for now running on a single PC): during each time step, each PID can execute $PID\_Speed_k$ operations.
By default, we set: $PID\_Speed_k = PID\_Speed = N/K$ (by default, PIDs are assumed to compute at the same speed).

When a PID is active, it increments \verb+count_active_k+ each time
an elementary operation (a diffusion from one node to another node in the same
$\Omega_k$ set, which roughly corresponds to a product of one entry $(F)_j$ with one entry
of the matrix $(P)_{ij}$ and the addition of the product to $(F)_i$) is done. 

Every time step, we set a local counter that counts the number
of elementary operations that are not consumed (because entering in the idle state).
When a PID  is idle, the {\em wasted} operations are then added to \verb+count_idle_k+.

In the following, the number of iterations is defined as the normalized quantity:
$$\frac{count\_active\_k + count\_idle\_k}{L}$$ 
so that it can be easily compared to the cost of one matrix-vector product,
or one iteration in power iteration.

\subsection{Computation cost}\label{sec:cost}
The computation effort of $PID_k$ is indirectly estimated through \verb+count_active_k+.
This counter is incremented:
\begin{itemize}
\item by one each time there is a local diffusion from one node to another;
\item by one to the receiver for each diffusion to one node (during fluid exchange) managed
  by the receiver; for the sender, we increment by one
  for each diffusion coming from $C_k(P) ([H]_k - [H_{old}]_k)$: this is the quantity that was
  underestimated in \cite{dist-test};
\item by the number of nodes re-affected for the partition set adaptation. 
\end{itemize}

\subsection{Partition sets}\label{sec:partition}
\subsubsection{Static partition sets}
As in \cite{dist-test}, we consider two simple $K$ partition sets for comparison purpose:
\begin{itemize}
\item Uniform partition: $\Omega_1 =\{1, 2, ..., N/K\}$,
  $\Omega_2 =\{N/K+1, 2, ..., 2\times N/K\}$, etc
\item Cost Balanced (CB) partition: $\Omega_k = \{\omega_k, \omega_k+1, ..., \omega_{k+1}-1\}$ such
  that $\sum_{n=\omega_k}^{\omega_{k+1}-1} (\#out_n) = L/K$,
\end{itemize}
such that $\{1,...,N\} = \Omega = \cup_k \Omega_k$.
The intuition of the cost balanced partition is that when we apply the diffusion iteration on
all nodes of each $\Omega_k$, the diffusion cost is constant.
The main reason why we chose this is the simplicity of its computation \cite{dist-test}. 

\subsubsection{Dynamic partition sets}
In the initial state, we start with the uniform or CB partition sets.
Then, we update the following quantity every time step ($PID\_Speed$ operations
in active or idle state):
\begin{eqnarray*}
&&\hspace*{-5mm}slope\_k :=\\
&&slope\_k \times (1-\eta) - log(r_k+s_k+\varepsilon)/log(10.0) \times \eta
\end{eqnarray*}
where $\varepsilon = target\_error/K/1000$ is added to avoid undefined value of $slope\_k$.
The quantity $-slope\_k$ measures the moving averaged value of the exponent
(base 10) of $r_k+s_k$: if we plot the curve $r_k+s_k$ as a function of the number
of iteration (normalized) in logscale on y-axis, the exponent represents the slope of the
curve. By default, we used $\eta = 0.5$.

Then, every time step, we compute $k$ which maximize and minimize
$slope\_k$ (resp. $i_{\max}$, $i_{\min}$). If the difference is more than 50\%:

\begin{eqnarray*}
\mbox{if } (slope\_min < slope\_max + \log(0.5)/\log(10.0))
\end{eqnarray*}

then, we reaffect:
$$
|\Omega_{i_{\min}}| \times \min\left(\frac{slope\_min+1}{slope\_max+1}, 0.1\right)
$$
nodes from $\Omega_{i_{\min}}$ to  $\Omega_{i_{\max}}$ ($i_{\min}$ identifies
the slowest PID).

To minimize the oscillation behaviour, the sets that are just re-affected (decreased or increased)
can not be re-affected during the next $Z$ steps (by default $Z=10$).

When the $\Omega_k$ set is re-affected, we increment its operation cost counter 
\verb+count_active_k+ (by the number of nodes modified for $\Omega_{i_{\min}}$ and  $\Omega_{i_{\max}}$).

\section{Experiments and evaluation}\label{sec:eval}

\subsection{Synthetic data} 
We first used a synthetic data generated as follows:
assuming a power-law $1/k^\alpha$ ($\alpha=1.5$ used here)
for the in-degree and the out-degree distribution,
we generated random links between pair of nodes (see \cite{dohy} for more details).

\subsubsection{Analysis of $K=2$: $N=1000$}
Let us start with 2 PIDs case for an easier illustration of the problem.
Figure \ref{fig:illustr-PID2-1000} shows the plots of the convergence speed
(given by the ratio of $r_k+s_k$ and the number of iterations) in logscale on y-axis, when
starting with a static partition sets of $250+750$, $500+500$ and $750+250$ (in this case,
$L=9543$).

\begin{figure}[htbp]
\centering
\includegraphics[angle=-90,width=\linewidth]{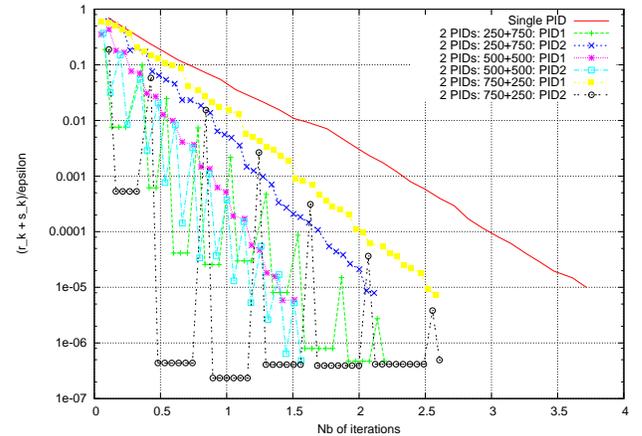}
\caption{Illustration of convergence speed: fluid exchange cost neglected.}
\label{fig:illustr-PID2-1000}
\end{figure}

When the $\Omega_k$ is not set correctly (too big or too small), the gain of
the parallelism is reduced. We remark that when the fluid exchange cost is neglected,
$K=2$ ($500+500$ case) can improve by factor above 2.
Figure \ref{fig:illustr-PID2-1000-cor} shows the plots of the convergence speed
integrating the fluid exchange cost: the relative gain to the single PID case
is much less important than previous results, illustrating the importance of this
factor even when $K$ is small.

\begin{figure}[htbp]
\centering
\includegraphics[angle=-90,width=\linewidth]{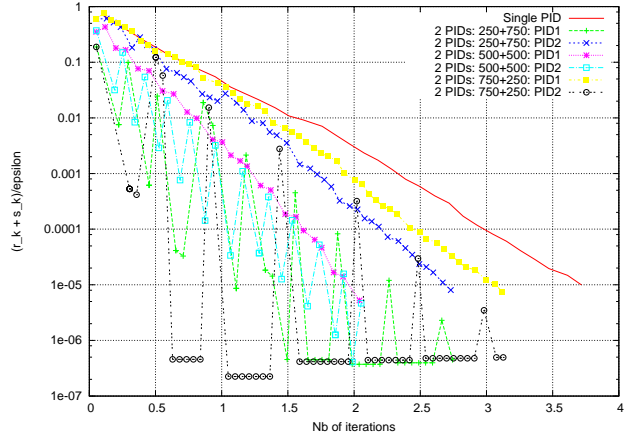}
\caption{Illustration of convergence speed: fluid exchange cost integrated.}
\label{fig:illustr-PID2-1000-cor}
\end{figure}

Figure \ref{fig:illustr-PID2-1000-adap} illustrates the impact of the partition
adaptation on the convergence speeds that are made closer: in this case, we took
initial partition sets of 750-250 and let the system self adapts.

\begin{figure}[htbp]
\centering
\includegraphics[angle=-90,width=\linewidth]{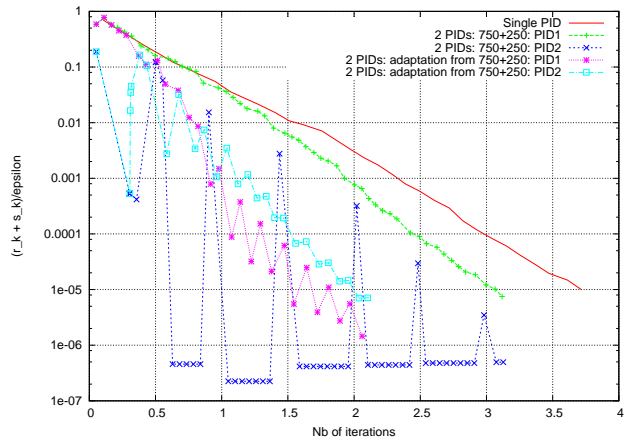}
\caption{Illustration of the impact of the dynamic partition.}
\label{fig:illustr-PID2-1000-adap}
\end{figure}

Figure \ref{fig:illustr-PID2-1000-adap-partition} illustrates the evolution of the partition
sets when starting from 750-250 (here, we took $Z=1$ for a quicker adaptation).

\begin{figure}[htbp]
\centering
\includegraphics[angle=-90,width=\linewidth]{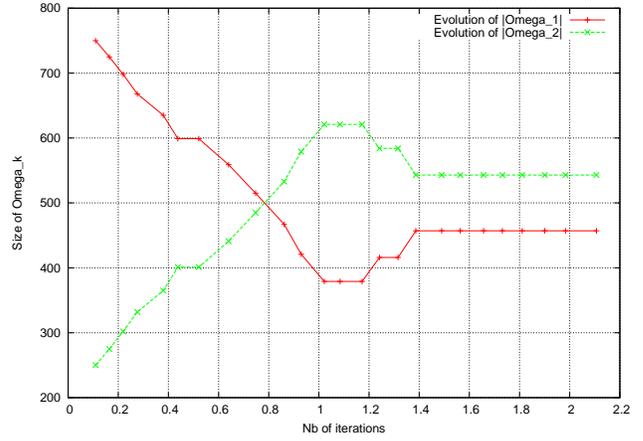}
\caption{Illustration of the evolution of the dynamic partition.}
\label{fig:illustr-PID2-1000-adap-partition}
\end{figure}

Table \ref{tab:compa1} gives a comparative computation time (number of iterations of the
slowest PID) of different approaches for $K=1$ to $128$ ($Z=10$): by construction, this can
be considered as the most favourable situation for the uniform partition (links are
independently and identically distributed to all nodes). We see that the dynamic strategy
can still improve in almost all situations (but not too much).

\begin{table}
\begin{center}
\begin{tabular}{|r|cc|cc|}
\hline
  & \multicolumn{2}{c|}{From Unif. partition} & \multicolumn{2}{c|}{From CB partition} \\
\hline
K & Static & Dynamic & Static & Dynamic\\
\hline
1 & 2.39 & 2.39 & 2.39 & 2.39\\
2 & 1.39 & 1.25 & 1.31 & 1.38\\
4 & 0.85 & 0.85 & 0.81 & 0.80\\
8 & 0.56 & 0.53 & 0.49 & 0.47\\
16 & 0.37 & 0.35 & 0.43 & 0.38\\
32 & 0.29 & 0.27 & 0.31 & 0.26\\
64 & 0.26 & 0.22 & 0.30 & 0.24\\
128 & 0.26 & 0.26 & 0.35 & 0.29\\
\hline
\end{tabular}\caption{Illustration of the computation time for a target error of $1/N$: $N=1000$.}\label{tab:compa1}
\end{center}
\end{table}

To further illustrate the advantage of the dynamic adaptation, we biased the nodes
ordering replacing the complete random one (previous) by the number of outgoing links (cf. Table \ref{tab:compa2}):
we see that the CB static strategy is not good and when $K\ge 4$ its performance is even degraded.

\begin{table}
\begin{center}
\begin{tabular}{|r|cc|cc|}
\hline
  & \multicolumn{2}{c|}{From Unif. partition} & \multicolumn{2}{c|}{From CB partition} \\
\hline
K & Static & Dynamic & Static & Dynamic\\
\hline
1 & 3.79 & 3.79 & 3.79 & 3.79\\
2 & 3.07 & 2.96 & 2.83 & 2.33\\
4 & 2.48 & 2.16 & 3.42 & 2.68\\
8 & 1.97 & 1.53 & 5.09 & 2.63\\
16 & 1.57 & 1.02 & 6.01 & 2.40\\
\hline
\end{tabular}\caption{Illustration of the computation time for a target error of $1/N$: $N=1000$. Nodes are ordered by the number of outgoing links.}\label{tab:compa2}
\end{center}
\end{table}

The results of the case when the nodes are ordered by the number of incoming links 
are shown in Table \ref{tab:compa3}: here the uniform partition is the worst one.

\begin{table}
\begin{center}
\begin{tabular}{|r|cc|cc|}
\hline
  & \multicolumn{2}{c|}{From Unif. partition} & \multicolumn{2}{c|}{From CB partition} \\
\hline
K & Static & Dynamic & Static & Dynamic\\
\hline
1 & 4.96 & 4.96 & 4.96 & 4.96\\
2 & 3.65 & 3.48 & 3.55 & 3.02\\
4 & 2.97 & 2.03 & 2.57 & 1.91\\
8 & 2.93 & 1.69 & 2.48 & 1.62\\
16 & 3.14 & 1.35 & 2.28 & 1.25\\
\hline
\end{tabular}\caption{Illustration of the computation time for a target error of $1/N$: $N=1000$. Nodes are ordered by the number of incoming links.}\label{tab:compa3}
\end{center}
\end{table}

Globally, what we observe is that when $N/K$ becomes too small, the gain is limited or the performance may be even
degraded due to the fluid exchange cost. Finally, we observe a very good
stability/performance of the dynamic partition strategy in all situations.

\subsection{Web graph datasets} 
For the evaluation purpose, we experimented the dynamic partition strategy on a
web graph imported from the dataset \verb+uk-2007-05@1000000+
(available on \cite{webgraphit}) which has
41,247,159 links on 1,000,000 nodes (45,766 dangling nodes).

Below we vary $N$ from 1,000 to 100,000 extracting from the dataset the
information on the first $N$ nodes.

\begin{table}
\begin{center}
\begin{tabular}{|l|ccc|}
\hline
N & L (nb links) & L/N & D (Nb dangling nodes)\\
\hline
1000 & 12,935 & 12.9 & 41 (4.1\%)\\
10000 & 125,439 & 12.5 & 80 (0.8\%)\\
100000 & 3,141,476 & 31.4 & 2729 (2.7\%)\\
\hline
\end{tabular}\caption{Extracted graph: $N=1000$ to $100000$.}
\end{center}
\end{table}

Figure \ref{fig:cost-compa-UP} shows the summarized results on the
convergence speeds (normalized to the convergence cost for $K=1$) for
$N=1000, 10000, 100000$ starting from the uniform partition
(unfortunately, we could not yet handle $N=1000000$ case because of the memory limitation
on a single PC).
We clearly see that because of the fluid exchange cost, the convergence becomes
slower when $K$ is too large compared to $N$ and that for larger $N$ the
optimal $K$ value is larger.

\begin{figure}[htbp]
\centering
\includegraphics[angle=-90,width=\linewidth]{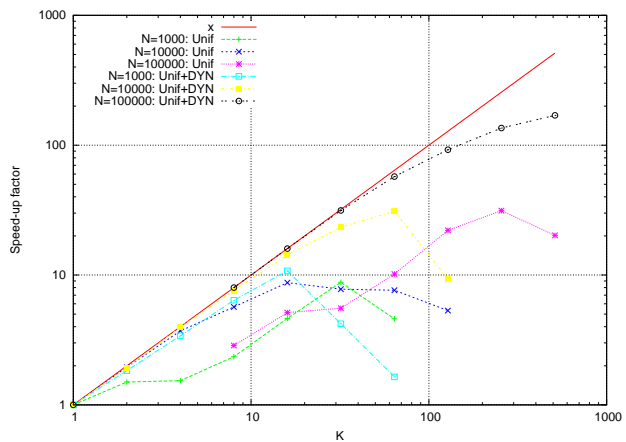}
\caption{Convergence speed-up factor: starting from uniform partition.}
\label{fig:cost-compa-UP}
\end{figure}

Figure \ref{fig:cost-compa-CB} shows the summarized results on the
convergence speeds starting from the CB partition. Figure \ref{fig:cost-compa-UP} and
Figure \ref{fig:cost-compa-CB} correct results reported in \cite{dist-test} 
(fluid exchange cost was underestimated) when $N/K$ is small.
However, the main conjecture/result which states that, when the computation is distributed 
over $K$ virtual machines, the computation speed increases almost linearly with $K$ 
with a slope becoming closer to one when the number $N$ of linear equations to be solved 
increases is still true: we conjecture that the slope goes to one asymptotically for large $N/K$ and
this is very clearly visible in the curves of the dynamic partition based approaches
(Unif$+$DYN or CB$+$DYN) when $N$ is increased.

\begin{figure}[htbp]
\centering
\includegraphics[angle=-90,width=\linewidth]{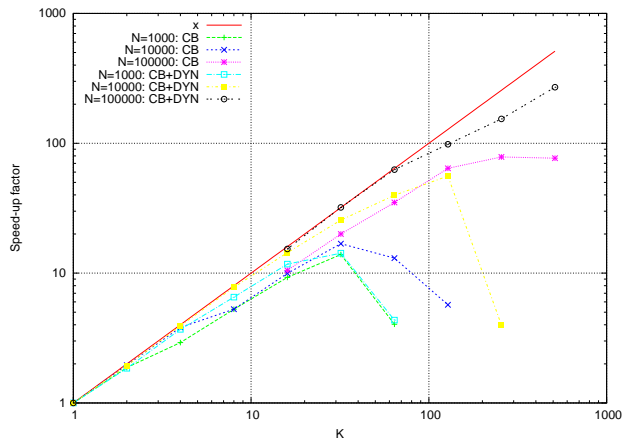}
\caption{Convergence speed-up factor: starting from CB partition.}
\label{fig:cost-compa-CB}
\end{figure}

Figure \ref{fig:idle-compa-10000} shows the consequence on the proportion of
the PIDs' idle state 
$$\frac{\sum_k count\_idle\_k}{\sum_k (count\_active\_k + count\_idle\_k)}$$ 
when different approaches are applied ($N=10000$). We see a clear reduction of the
idle state with the dynamic strategy when the fluid exchange is not dominant. 

\begin{figure}[htbp]
\centering
\includegraphics[angle=-90,width=\linewidth]{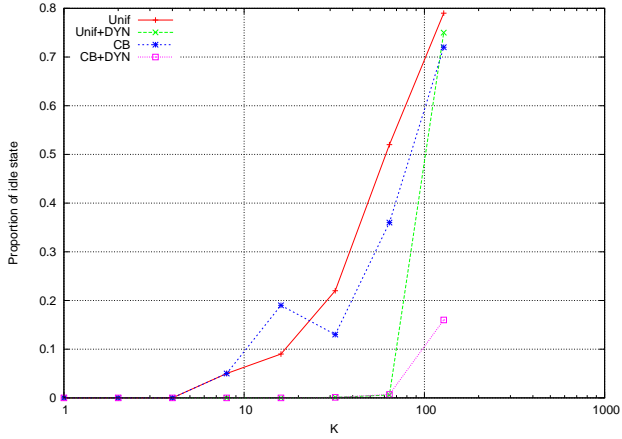}
\caption{Proportion of the idle state: $N=10000$.}
\label{fig:idle-compa-10000}
\end{figure}

Figure \ref{fig:evol-cv-K2} shows the typical result of two different
convergence speeds: in this case PID2 is the slowest one. The fluid exchange is done every
about 1.2 iterations which is clearly visible here.
We can see that PID1 can enter in the idle state because it is waiting for inputs
from PID2 (for $x$ between 4 and 5, between 6.5 and 7.5 etc) when it reaches the target
value $\max(s_k/10.0, target\_error\times\epsilon/K/10)$: this is globally not optimal
in terms of the PID's computation capacity utilization. 

\begin{figure}[htbp]
\centering
\includegraphics[angle=-90,width=\linewidth]{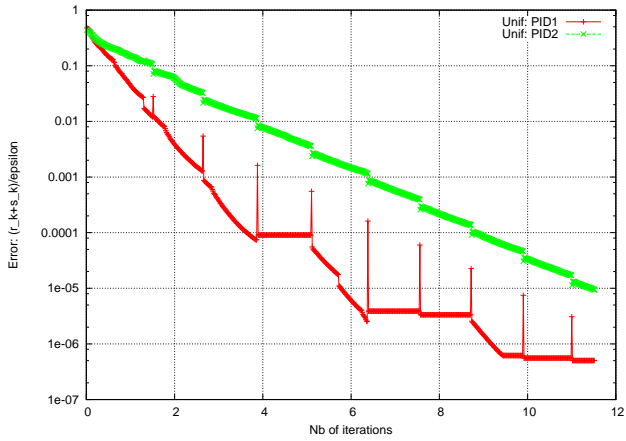}
\caption{Evolution of convergence: $K=2$, $N=100000$ with static uniform partition.}
\label{fig:evol-cv-K2}
\end{figure}

Figure \ref{fig:evol-cv-K2-DYN} shows the impact of the dynamic partition starting
from the uniform partition for the same case than Figure \ref{fig:evol-cv-K2}.
The corresponding evolution of the partition sets is shown in Figure \ref{fig:evol-cv-K2-DYN-omega}.

\begin{figure}[htbp]
\centering
\includegraphics[angle=-90,width=\linewidth]{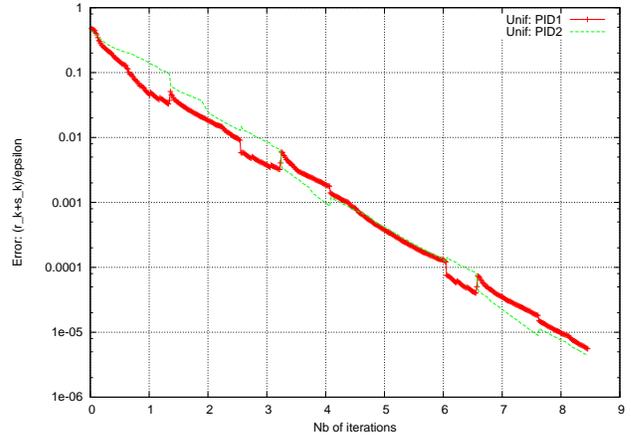}
\caption{Evolution of convergence: $K=2$, $N=100000$. Dynamic partition from the uniform partition.}
\label{fig:evol-cv-K2-DYN}
\end{figure}

\begin{figure}[htbp]
\centering
\includegraphics[angle=-90,width=\linewidth]{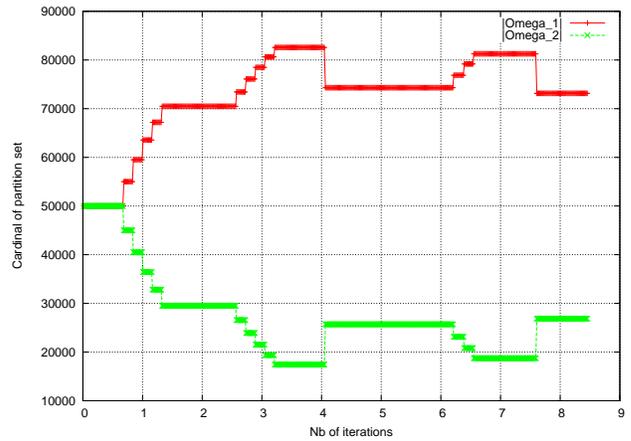}
\caption{Evolution of partition sets: $K=2$, $N=100000$.}
\label{fig:evol-cv-K2-DYN-omega}
\end{figure}

Figure \ref{fig:evol-cv-K2-CB} shows the evolution of the convergence of PIDs
with static CB partition: because CB is based on a heuristic simplification, it does
not guarantee the same computation effort for the two PIDs.

\begin{figure}[htbp]
\centering
\includegraphics[angle=-90,width=\linewidth]{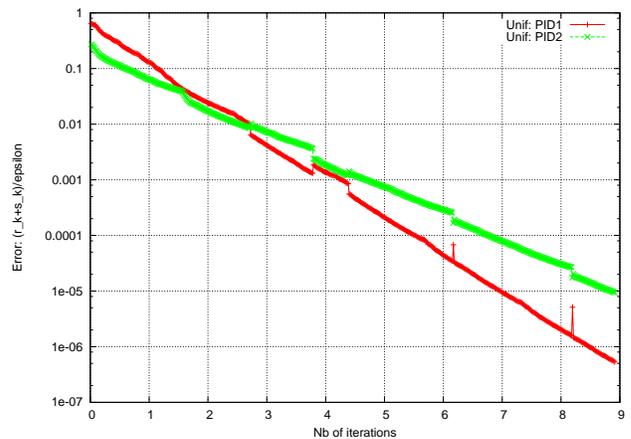}
\caption{Evolution of convergence: $K=2$, $N=100000$ with static CB partition.}
\label{fig:evol-cv-K2-CB}
\end{figure}

Figure \ref{fig:evol-cv-K4} shows the evolution of the convergence with $K=4$
with the static uniform partition, the static CB partition and the dynamic partition
starting from the uniform and CB partitions:
in this case, the benefit of the dynamic partition is very clear with an acceleration
by a factor above 3.

\begin{figure}[htbp]
\centering
\includegraphics[angle=-90,width=\linewidth]{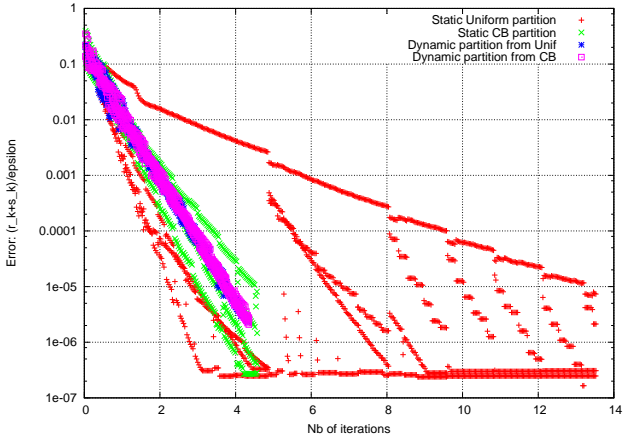}
\caption{Evolution of convergence: $K=4$, $N=100000$. Comparison of static unif., static CB and dynamic from unif. and CB.}
\label{fig:evol-cv-K4}
\end{figure}

Figure \ref{fig:evol-cv-K8} shows the evolution of the convergence with $K=8$
with the static uniform partition, the static CB partition and the dynamic partition
starting from the uniform partition:
in this case, the speed-up factor with the dynamic strategy is above 2.

\begin{figure}[htbp]
\centering
\includegraphics[angle=-90,width=\linewidth]{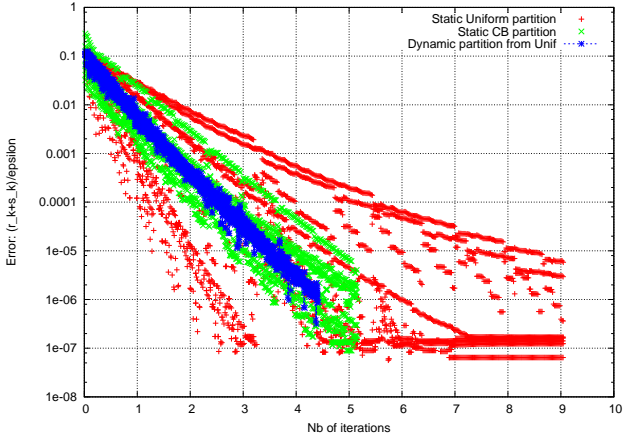}
\caption{Evolution of convergence: $K=8$, $N=100000$. Comparison of static unif., static CB and dynamic from unif.}
\label{fig:evol-cv-K8}
\end{figure}

Figure \ref{fig:evol-cv-K128} shows the result of 128 different
convergence speeds with $K=128$: in this case, we can identify 2 slowest PIDs.
The computation capacities of 126 other PIDs are likely to be wasted.

\begin{figure}[htbp]
\centering
\includegraphics[angle=-90,width=\linewidth]{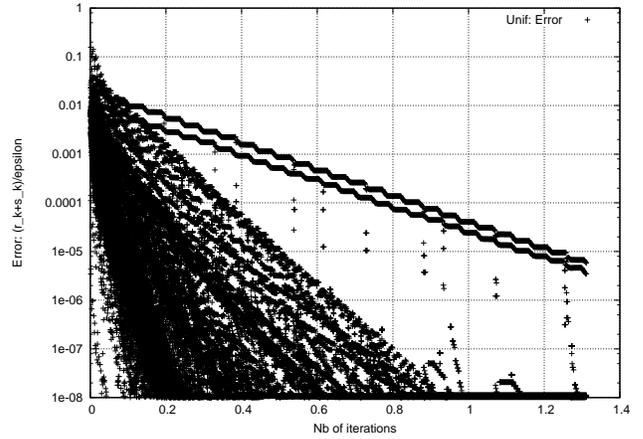}
\caption{Evolution of convergence: $K=128$, $N=100000$ with static uniform partition.}
\label{fig:evol-cv-K128}
\end{figure}

Figure \ref{fig:evol-cv-K128-DYN} shows the impact of the dynamic partition starting
from the uniform partition for the same case than Figure \ref{fig:evol-cv-K128}.
In this case, the speed-up factor is about 4 thanks to a better computation effort
redistribution between PIDs.

\begin{figure}[htbp]
\centering
\includegraphics[angle=-90,width=\linewidth]{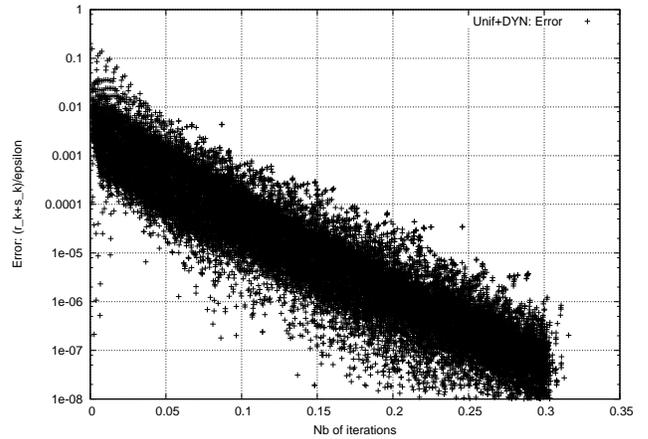}
\caption{Evolution of convergence: $K=128$, $N=100000$. Dynamic partition from the uniform partition.}
\label{fig:evol-cv-K128-DYN}
\end{figure}

Figure \ref{fig:global-cv-10000} and Figure \ref{fig:global-cv-10000-b} show the 
global convergence (an upper bound on the $L_1$ norm to the distance) for different approaches ($N=10000$):
the benefit of the dynamic adaptation is more visible for $K\ge 8$.
Note that those curves must be strictly decreasing function: we observe here some local fluctuation
due to the artefact of the time stepped approximation (linked to the fluid exchange cost): when
the fluid exchange cost becomes important, the concerned PID is likely to be frozen during that time.

\begin{figure}[htbp]
\centering
\includegraphics[angle=-90,width=\linewidth]{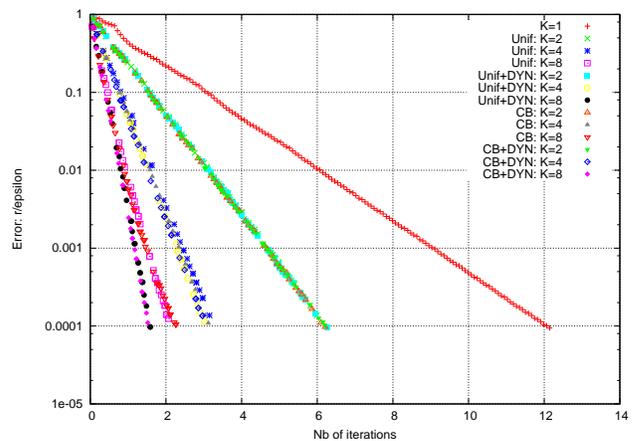}
\caption{Global convergence: $N=10000$. For $K=2,4,8$.}
\label{fig:global-cv-10000}
\end{figure}

\begin{figure}[htbp]
\centering
\includegraphics[angle=-90,width=\linewidth]{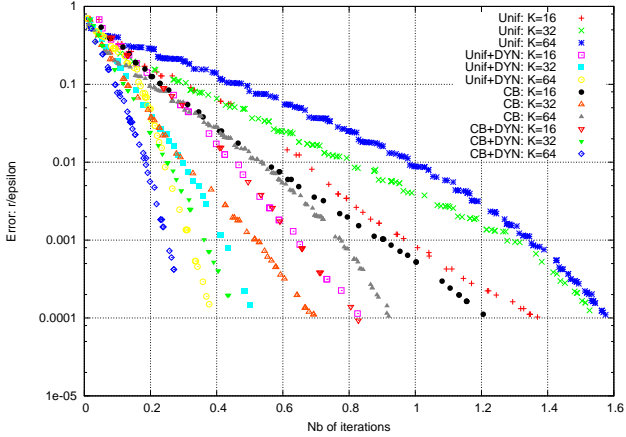}
\caption{Global convergence: $N=10000$. For $K=16,32,64$.}
\label{fig:global-cv-10000-b}
\end{figure}

Figure \ref{fig:global-cv-100000-b} and Figure \ref{fig:global-cv-100000-c} show the 
global convergence for $N=100000$:
when $K$ and $N$ are larger, the analysis becomes much more complex: we can observe 
significant and sudden slope modification during the iteration. See for instance
Unif$+$DYN or CB curves for $K=512$ in Figure \ref{fig:global-cv-100000-c}.
One of very visible effect is the impact of the fluid exchange cost which is 
increased for larger value of $K$.

\begin{figure}[htbp]
\centering
\includegraphics[angle=-90,width=\linewidth]{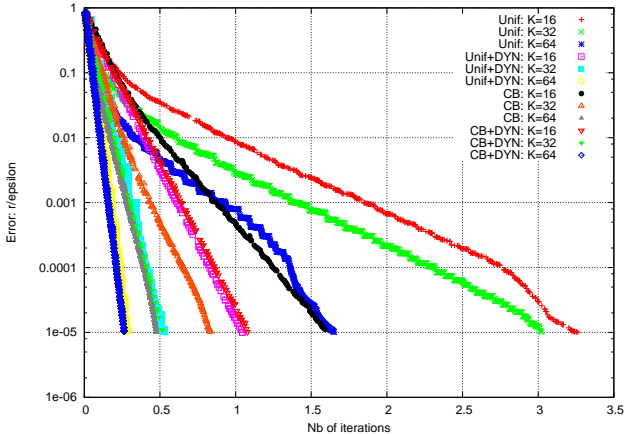}
\caption{Global convergence: $N=100000$. For $K=16,32,64$.}
\label{fig:global-cv-100000-b}
\end{figure}

\begin{figure}[htbp]
\centering
\includegraphics[angle=-90,width=\linewidth]{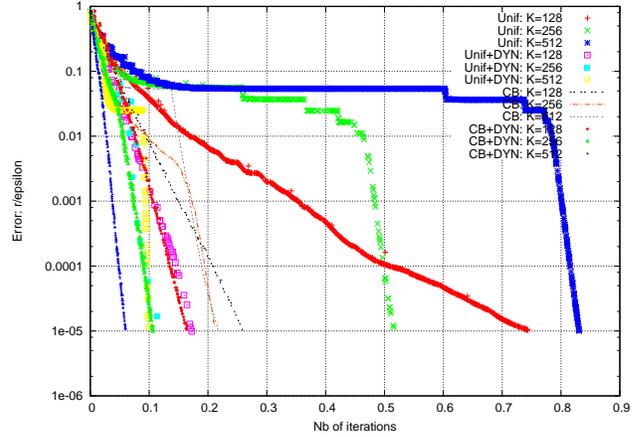}
\caption{Global convergence: $N=100000$. For $K=128,256,512$.}
\label{fig:global-cv-100000-c}
\end{figure}

The above results show in particular (and this is not surprising)
that it is not necessarily better to increase $K$ and
an optimal $K$ need to be applied (for a given vector size $N$).
This may suggest the possibility of considering a further adaptive scheme where we could
also dynamically adjust the number of PIDs: we hope to address this issue in a future
work.
What we propose here is a first simple candidate to highlight the potential of
the approach. From this first step, one may explore
a lot of variants (for instance, we should favour partition sets such that there are more
links inside the $\Omega_k$ sets; we could also define the number of nodes to be
re-affected, when modification required, based on its CB evaluation, etc).

\section{Conclusion}\label{sec:conclusion}
In this paper, we presented an adaptive dynamic partition strategy 
applied to a distributed computation architecture of the D-iteration method.
Through experiments on synthetic data and real dataset, we showed that a dynamic partition
strategy brings a robustness and a better efficiency guarantee compared to the static partition
strategy, especially when $N$ is large.
We believe that, even though this is preliminary results that need to be confirmed by a real
deployment of a distributed system with possibly further adaptation/modification of the
algorithm design, we showed here the potential of a new promising distributed computation architecture
to solve a very large diagonal dominant class of linear systems.


\end{psfrags}
\bibliographystyle{abbrv}
\bibliography{sigproc}

\begin{thebibliography}{10}

\bibitem{webgraphit}
http://law.dsi.unimi.it/datasets.php.

\bibitem{serge}
S.~Abiteboul, M.~Preda, and G.~Cobena.
\newblock Adaptive on-line page importance computation.
\newblock {\em WWW2003}, pages 280--290, 2003.

\bibitem{Arasu02pagerankcomputation}
A.~Arasu, J.~Novak, J.~Tomlin, and J.~Tomlin.
\newblock Pagerank computation and the structure of the web: Experiments and
  algorithms, 2002.

\bibitem{Bagnara95aunified}
R.~Bagnara.
\newblock A unified proof for the convergence of jacobi and gauss-seidel
  methods.
\newblock {\em SIAM Review}, 37, 1995.

\bibitem{Bertsekas:1989:PDC:59912}
D.~P. Bertsekas and J.~N. Tsitsiklis.
\newblock {\em Parallel and distributed computation: numerical methods}.
\newblock Prentice-Hall, Inc., Upper Saddle River, NJ, USA, 1989.

\bibitem{bian}
M.~Bianchini, M.~Gori, and F.~Scarselli.
\newblock Inside pagerank.
\newblock {\em ACM Trans. Internet Techn.}, 2005.

\bibitem{Boldi2009}
P.~Boldi, M.~Santini, and S.~Vigna.
\newblock Pagerank: Functional dependencies.
\newblock {\em ACM Trans. Inf. Syst.}, 27:19:1--19:23, November 2009.

\bibitem{brezinski}
C.~Brezinski, M.~Redivo-Zaglia, and S.~Serra-Capizzano.
\newblock Extrapolation methods for pagerank computations.
\newblock {\em Comptes Rendus Acad. Sci.}, pages 393--397, 2005.

\bibitem{Bunch:1989:SSA:75554.75560}
J.~R. Bunch, J.~W. Demmel, and C.~F. van Loan.
\newblock The strong stability of algorithms for solving symmetric linear
  systems.
\newblock {\em SIAM J. Matrix Anal. Appl.}, 10:494--499, October 1989.

\bibitem{francis}
J.~G.~F. Francis.
\newblock The {QR} transformation, {I}.
\newblock {\em The Computer Journal}, 4(3):265--271, 1961.

\bibitem{Golub1996}
G.~H. Golub and C.~F.~V. Loan.
\newblock {\em Matrix Computations}.
\newblock The Johns Hopkins University Press, 3rd edition, 1996.

\bibitem{haveliwala}
T.~H. Haveliwala, S.~D. Kamvar, D.~Klein, C.~D. Manning, and G.~H. Golub.
\newblock Computing pagerank using power extrapolation.
\newblock {\em Technical report, Stanford University}, 2003.

\bibitem{distributed}
D.~Hong.
\newblock D-iteration based asynchronous distributed computation.
\newblock {\em arXiv, http://arxiv.org/abs/1202.3108}, February 2012.

\bibitem{dist-test}
D.~Hong.
\newblock D-iteration: Evaluation of the asynchronous distributed computation.
\newblock {\em submitted, http://arxiv.org/abs/1202.6168}, February 2012.

\bibitem{update}
D.~Hong.
\newblock D-iteration: Evaluation of the update algorithm.
\newblock {\em arXiv, http://arxiv.org/abs/1202.6136}, February 2012.

\bibitem{d-algo}
D.~Hong.
\newblock D-iteration method or how to improve gauss-seidel method.
\newblock {\em arXiv, http://arxiv.org/abs/1202.1163}, February 2012.

\bibitem{dohy}
D.~Hong.
\newblock Optimized on-line computation of pagerank algorithm.
\newblock {\em submitted, http://arxiv.org/abs/1202.6158}, 2012.

\bibitem{kirkland}
I.~C.~F. Ipsen and S.~Kirkland.
\newblock Convergence analysis of a pagerank updating algorithm by langville
  and meyer.
\newblock {\em SIAM J. Matrix Anal. Appl.}, 27(4):952--967, 2006.

\bibitem{jela}
M.~Jelasity, G.~Canright, and K.~Eng\o-Monsen.
\newblock Asynchronous distributed power iteration with gossip-based
  normalization.
\newblock In {\em Euro-Par 2007 Parallel Processing}, volume 4641 of {\em
  Lecture Notes in Computer Science}, pages 514--525. 2007.

\bibitem{kamvar}
S.~D. Kamvar, T.~H. Haveliwala, and G.~H. Golub.
\newblock Adaptive methods for the computation of pagerank.
\newblock {\em Linear Algebra Appl.}, pages 51--65, 2004.

\bibitem{kamvar2}
S.~D. Kamvar, T.~H. Haveliwala, C.~D. Manning, and G.~H. Golub.
\newblock Extrapolation methods for accelerating pagerank computations.
\newblock {\em Proc. of International World Wide Web Conference (WWW2003)},
  pages 261--270, 2003.

\bibitem{Kohlschutter06efficientparallel}
C.~Kohlsch\"utter, P.-A. Chirita, R.~Chirita, and W.~Nejdl.
\newblock Efficient parallel computation of pagerank.
\newblock In {\em In Proc. of the 28th European Conference on Information
  Retrieval}, pages 241--252, 2006.

\bibitem{DBLP:journals/corr/abs-cs-0606047}
G.~Kollias, E.~Gallopoulos, and D.~B. Szyld.
\newblock Asynchronous iterative computations with web information retrieval
  structures: The pagerank case.
\newblock {\em CoRR}, abs/cs/0606047, 2006.

\bibitem{Koutis10approachingoptimality}
I.~Koutis, G.~L. Miller, and R.~Peng.
\newblock Approaching optimality for solving sdd linear systems, 2010.

\bibitem{kub}
V.~N. Kublanovskaya.
\newblock On some algorithms for the solution of the complete eigenvalue
  problem.
\newblock {\em USSR Computational Mathematics and Mathematical Physics},
  3:637--657, 1961.

\bibitem{deep}
A.~N. Langville and C.~D. Meyer.
\newblock Deeper inside pagerank.
\newblock {\em Internet Mathematics}, 1(3), 2004.

\bibitem{lang2}
A.~N. Langville and C.~D. Meyer.
\newblock Updating the stationary vector of an irreducible markov chain with an
  eye on google's pagerank.
\newblock {\em Technical report, North Carolina State University}, 2004.

\bibitem{Lubachevsky:1986:CAA:4904.4801}
B.~Lubachevsky and D.~Mitra.
\newblock A chaotic asynchronous algorithm for computing the fixed point of a
  nonnegative matrix of unit spectral radius.
\newblock {\em J. ACM}, 33(1):130--150, Jan. 1986.

\bibitem{marek}
I.~Marek and P.~Mayer.
\newblock Convergence theory of some classes of iterative
  aggregation/disaggregation methods for computing stationary probability
  vectors of stochastic matrices.
\newblock {\em Linear Algebra Appl.}, pages 177--200, 2003.

\bibitem{page}
L.~Page, S.~Brin, R.~Motwani, and T.~Winograd.
\newblock The pagerank citation ranking: Bringing order to the web.
\newblock {\em Technical Report Stanford University}, 1998.

\bibitem{Saad}
Y.~Saad.
\newblock {\em Iterative Methods for Sparse Linear Systems}.
\newblock Society for Industrial and Applied Mathematics, Philadelphia, PA,
  USA, 2nd edition, 2003.

\bibitem{DBLP:journals/corr/cs-DS-0310036}
D.~A. Spielman and S.-H. Teng.
\newblock Solving sparse, symmetric, diagonally-dominant linear systems in time
  o(m$^{\mbox{1.31}}$).
\newblock {\em CoRR}, cs.DS/0310036, 2003.

\bibitem{mises}
R.~von Mises and H.~Pollaczek-Geiringer.
\newblock Praktische verfahren der {Gleichungsaufl\"osung}.
\newblock {\em ZAMM - Zeitschrift {f\"{u}r} Angewandte Mathematik und
  Mechanik}, 9:152--164, 1929.

\end{thebibliography}

\end{document}